\documentclass[iop]{emulateapj-rtx4}
\usepackage{apjfonts}

\newcommand{\Msun}{M$_{\odot}$}

\shorttitle{Constraints on NGC\,6397 age}
\shortauthors{Correnti et al.}

\begin{document}


\title{The age of the old metal-poor Globular Cluster NGC\,6397 using WFC3/IR photometry\altaffilmark{1}} 


\author{Matteo Correnti\altaffilmark{2}, 
Mario Gennaro\altaffilmark{2}, Jason S. Kalirai\altaffilmark{2,3}, Roger E. Cohen\altaffilmark{2}, and Thomas M. Brown\altaffilmark{2}}

\altaffiltext{1}{Based on observations with the NASA/ESA {\it Hubble Space Telescope}, obtained at the Space Telescope Science Institute, which is operated by the Association of Universities for Research in Astronomy, Inc., under NASA contract NAS5-26555}

\altaffiltext{2}{Space Telescope Science Institute, 3700 San Martin Drive, Baltimore, MD 21218, USA; correnti, jkalirai, gennaro, tbrown, calamida@stsci.edu}

\altaffiltext{3}{Center for Astrophysical Science, John Hopkins University, Baltimore, MD 21218, USA}

\begin{abstract}
{Globular Clusters (GCs) in the Milky Way represent the ideal laboratory to establish the age of the oldest stellar populations and to measure the color-magnitude relation of stars. Infrared (IR) photometry of these objects provides a new opportunity to accomplish this task. In particular, at low stellar masses, the stellar main sequence (MS) in an IR color-magnitude diagram (CMD) exhibits a sharp ``kink'' (due to opacity effects in M dwarfs), such that lower mass and cooler dwarfs become bluer in the $F110W - F160W$ color baseline and not redder. This inversion of the color-magnitude relation offers the possibility to fit GC properties using IR imaging, and to reduce their uncertainties. Here, we used the IR channel of the Wide Field Camera 3 onboard the {\it Hubble Space Telescope} to obtain new, deep high-resolution photometry of the old metal-poor GC NGC\,6397. From the analysis of the GC CMD, we revealed below the MS ``kink'' the presence of two MSs with different chemical composition. We derived the cluster fiducial line and we compared it with a grid of isochrones over a large range of parameter space, allowing age, metallicity, distance and reddening to vary freely within reasonable selected ranges. We derived an age of 12.6 Gyr with a random uncertainty $\sigma \sim$ 0.7 Gyr. These results confirm that the analysis of the IR color-magnitude of stars provide a valuable tool to measure the GC ages and offers a new venue to determine their absolute age to sub-Gyr accuracy with next generation IR telescopes.}\end{abstract}





\section{Introduction}
\label{s:intro}

Globular Clusters (GCs) are among the oldest objects in the Universe for which accurate ages can be determined. Determining the absolute age of these objects is a measure of when baryonic structure formation occurred in the Universe \citep{sper+03}, thus providing a strict lower limit on its age. Moreover, it provides robust constraints on the physics adopted in stellar evolutionary models \citep{salwei98,cass+99,vand+08,dott+08}. Measuring the relative age difference between clusters associated with distinct structural components of the Galaxy also yields the formation and assembly timescale of these parent populations. 

Derivation of star cluster ages has primarily relied on the analysis of visible-light color-magnitude diagrams (CMDs) and comparison with stellar evolutionary models. However, uncertainties in the other fundamental parameters have hampered the estimates of absolute GC ages. In addition to the systematic uncertainties related to the different assumption in the stellar models \citep[$\sigma$ = 0.4 Gyr \,--\,][]{chakra02,vall+13a,vall+13b} and the cluster metallicity \citep[$\sigma$ = 0.5 Gyr for a 0.2 dex error \,--\,][]{dott+08} the largest uncertainty impacting this technique comes from simultaneously ``fitting'' the age at a given distance and reddening \citep[$\sigma$ = 1.5\,--\,2 Gyr \,--\,][]{chab08}. 

Recently, a new tool was introduced to measure accurate star cluster parameters based on high precision infrared (IR) CMDs. \citet{sara+09} and \citet{bono+10} demonstrated that the MS in a pure IR CMD exhibits a change in the color-magnitude relation of stars at $\sim$ 0.5 \Msun, with color rapidly shifting to the blue \citep{pulo+98,pulo+99,zocc+00,cala+09,kali+12,milo+12b,milo+14,mone+15}. We will refer to this feature as the MS kink (hereafter MSK). This feature arises from a redistribution of the emerging stellar flux due to a change in opacity caused by the collision-induced absorption of H$_2$ \citep[][and references therein]{lins69,saum+94}. This feature presents two fundamental characteristics: the MSK luminosity and the bending shape are dependent on metallicity, but are independent of age beyond $\sim$ 1 Gyr, allowing to partially break the degeneracy between these two parameters. Furthermore, the shape of the color-magnitude relation, in particular its shift to the blue at magnitudes fainter than the MSK, provides a new opportunity to simultaneously constrain the distance and reddening accurately.   

In this context, in \citet[][hereafter C16]{corr+16} we demonstrated the predictive power and the constraints imposed on GC properties by the observation of this feature. We analyzed {\it Hubble Space Telescope (HST)} Wide Field Camera 3 IR (WFC3/IR) archival observations of four GCs  (namely, 47\,Tuc, M\,4, NGC\,2808, NGC\,6752) for which the data are deep enough to reach at least $\simeq$ 2 mag below the MSK, allowing us to fully sample the bending of the MS at low masses. From the comparison of the fiducial lines, obtained with an ad-hoc fitting method, with a grid of isochrones, we derived the best-fit parameters, we quantified the correlations among them and we obtained the individual uncertainties. From our analysis, we derived random age uncertainties of $\sigma \sim$ 0.7\,--\,1.1 Gyr. Hence, observing the near-IR MSK offers a new venue to push the absolute age of GCs to sub-Gyr statistical accuracy. 

Although the clusters analyzed in C16 provide a fairly good sample to test the predictive power of the MSK, they do not cover the entire metallicity regime probed by Galactic GCs, spanning only a limited range ([Fe/H] $\simeq$ -0.7\,--\,-1.5 dex). Unfortunately, our sample lacked a genuine old metal-poor cluster. Age measurements for these old metal-poor stellar system is of fundamental importance. For example, if it can be established that these objects formed $>$ 13 Gyr ago, then they must have formed in very low mass halos and they have been affected by reionization \citep{buljoh05,gned10}.

In this context, we obtained new {\it HST} WFC3/IR observations of the old metal-poor GC NGC\,6397. This cluster represents an ideal candidate to complement our sample, being near \citep[d $\sim$ 2.3 kpc,][]{harr06} and one of the most metal-poor cluster in the Milky Way ([Fe/H] $\simeq$ -2.0 dex, Gratton et al. 03). Performing the same analysis as in C16, we derived the GC best-fit parameters and their uncertainties. This study allows us to obtain an independent and statistical accurate age for the GC NGC\,6397. Moreover, adding NGC\,6397 to the C16 sample provides a unique opportunity to test the age-metallicity relation of Milky Way GCs. In fact, NGC\,6397 data provide a point to anchor the age-metallicity relation in the low metallicity regime. 

The remainder of the paper is organized as follows: the data set and reduction is presented in Section~2. In Section~3 we show the cluster CMD and Section~4 describe the method adopted to obtain its fiducial line. In Section~5 we derived the best-fit isochrones and the probability distribution functions, thus obtaining the cluster best-fit parameters and their uncertainties. In Section~6 we show the age-metallicity relation obtained from the whole sample (NGC\,6397 and the 4 GCs analyzed in C16). Finally, in Section~7 we summarize and discuss the results. 

\section{Observations and Data Reduction}
\label{s:reduction}
NGC\,6397 was observed with {\it HST} on 2016 March 08 using the IR channel of the WFC3 as part of the {\it HST} program 14124 (PI: M. Correnti). The cluster was observed in the two filters {\em F110W} and {\em F160W}. Four exposures were taken in each filter, with exposure times of $\sim$ 250 s and $\sim$ 400 s, respectively. To resample the PSF, mitigate hot pixels, and minimize errors from flat fielding we used a 4-point dither pattern in each filter. To ensure that the confusion limit does not truncate the photometric depth reached in the cluster, observations were taken at approximately 1\farcm3 from NGC\,6397's center. 

To reduce NGC\,6397 images we used the following approach: calibrated \texttt{.flt} science images and a drizzled, stacked, distortion-corrected \texttt{.drz} reference image were retrieved form the {\it HST} archive. The first are the products of the \texttt{calwf3} data reduction pipeline. The latter are the results of the combination of the \texttt{.flt} files using {\it AstroDrizzle} with a default parameter set, as provided by the {\it WFC3} Team. The pipeline uses the latest reference files to perform bad pixel flagging and corrections for dark current, non-linearity, flat fielding and up-the-ramp fitting of multiple readouts per WFC3/IR exposure. Additional preprocessing and PSF photometry were performed using the latest version of \texttt{Dolphot}\footnote{\url{http://americano.dolphinism.com}} \citep[][and numerous subsequent updates]{dolp00}, which includes pre-processing and photometry modules customized to each camera onboard {\it HST} as well as pre-computed PSFs customized to each filter. Before performing PSF photometry, the WFC3/IR module of \texttt{Dolphot} was used to mask bad pixels and generate a sky frame corresponding to each science image\footnote{We set \texttt{step} = 2 to generate high-resolution sky frames}. Iterative PSF photometry was performed simultaneously on the \texttt{.flt} images using the drizzled \texttt{.drz} image as a reference frame for alignment. \texttt{Dolphot} has many parameters governing the details of the image alignment, photometry and aperture corrections, and a detailed description can be found in \citet{will+14}. After testing on the NGC\,6397 data as well as sets of similar WFC/IR imaging of Galactic bulge GCs (Cohen et al., in prep.), using both observed catalogs and \texttt{Dolphot} artificial star tests, we found that the optimal compromise between photometric completeness and ability to reject spurious sources resulted from a hybrid between the parameters used by \citet{will+14} and those recommended by the \texttt{Dolphot} manual. 

Specifically, the parameters modified from their default values were \texttt{Force} = 1 (which forces all sources to be classified as stars, improving precision in crowded fields at cost of background galaxy discrimination), \texttt{SigFind} = 3 , \texttt{PosStep} = 0.1, \texttt{RCentroid} = 1, and \texttt{RCombine} = 1.415. With these settings we achieved optimal image alignment and source detection.

The resulting photometric catalogs contain photometry in the Vegamag system as well as several diagnostic parameters (including sharpness, roundness, and crowding) which are useful to cull remaining artifacts and spurious detections. Based on the aforementioned tests, we found that requiring $\vert$\texttt{sharp}$\vert$ $\leq$ 0.1, \texttt{crowd} $\leq$ 0.25, \texttt{round} $\leq$ 0.4 and a photometric quality flag of 2 or less in each filter efficiently eliminated spurious detections while minimizing the loss of well-measured stellar sources. \citep[e.g.][]{cohen+14, dieb+16}. Artificial star tests reveal that our photometry has S/N $>$ 100 and a (spatially integrated) completeness of $>$ 50\% down to $F160W \sim$ 20, nearly 2 mag faintward of the MS kink \citep[for a detailed description of artificial star tests, see][]{cohen+18}. 

\section{Color-Magnitude Diagram: a MS split}
\label{s:cmd}

\begin{figure}[!thp]
\includegraphics[width=1\columnwidth]{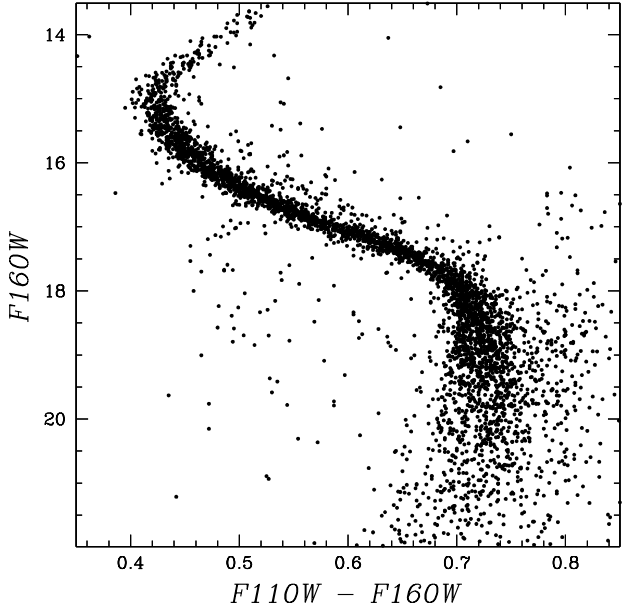}
\caption{$F160W$ vs $F160W - F110W$ NGC\,6397 CMD. $F160W$ vs $F110W - F160W$ NIR CMD of NGC\,6397 exhibits a rich main sequence extending from the subgiant branch down to very low mass M dwarfs.  The distinctive MSK feature at $F160W$ $\sim$ 18 mag is easily measured.}
\label{f:cmd}
\end{figure}

Figure~\ref{f:cmd} shows the $F160W$ vs $F110W - F160W$ CMD for NGC\,6397. A visual inspection of the cluster CMD shows that above the MSK the MS is narrow and there are no evident signs of any large color spread or split. Conversely, at fainter luminosity below the MSK, the MS broadens and it seems to be separated in two sequences. The presence of a double MS in NGC\,6397 was already reported in \citet{milo+12a}. From the analysis of multi-band {\it HST} ACS/WFC and WFC3/UVIS photometry, the authors found that the cluster MS splits in two components: a primordial population, containing $\sim$ 30\% of the stars, and a second generation with enhanced sodium and nitrogen, depleted carbon and oxygen, and a slightly enhanced helium abundance ($\Delta Y \sim$ 0.01).

The presence of a MS split below the MSK in IR CMD is not completely unexpected: as discovered by \citet{milo+12b,milo+14} and confirmed in C16, for the GCs NGC\,2808 and M\,4, multiple populations, observed at larger masses, in optical bandpasses, are visible also in the IR at low stellar masses, below the MSK. The presence of a double MS, below the MKS, is due to a different chemical composition of the two sub-populations, with the blue MS representing the first stellar generation, having primordial helium and high oxygen, and the redder MS associated with a second generation consisting of stars enhanced in helium, nitrogen, sodium, and depleted in oxygen. Recently, \citet{milo+17} demonstrated that the morphology of the IR MS below the MSK in the GC $\omega$ Cen is even more complex, indicating the presence of multiple sub-populations. 

\begin{figure}[!thp]
\includegraphics[width=1\columnwidth]{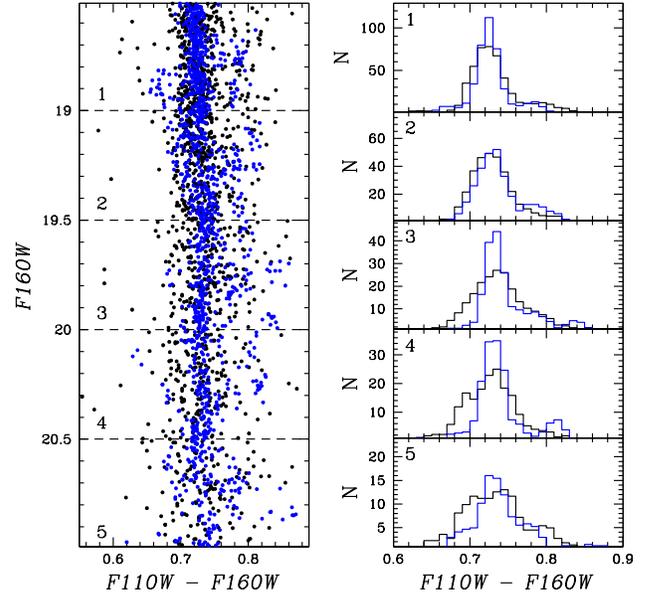}
\caption{Left panel: a zoom-in view of the NGC\,6397 CMD in the region below the MSK, with superimposed the simulated CMD (blue dots) obtained as described in Sect.~\ref{s:cmd} . Right panel: color distribution of observed and simulated stars (black and blue lines, respectively) in the five magnitude intervals indicated by the dashed lines in the left panel. The presence of a double MS become evident at magnitude fainter than $F160W <$ 19.5 mag.}
\label{f:cmdbin1}
\end{figure}

To verify the significance of the MS split suggested by the visual inspection of the CMD, we derived the color distribution of stars, in fixed magnitude intervals, in the region below the MSK. The left panel of Figure~\ref{f:cmdbin1} shows a zoom-in view of the NGC\,6397 CMD, whereas the color distribution of stars in the five magnitude intervals, over the range 18.50 $< F160W <$ 21.00 mag, are shown in the right panel. Figure~\ref{f:cmdbin1} shows a progressive broadening of the MS as a function of the magnitude, in particular below $F160W >$ 19.50 mag, where the two sequences form two distinct peaks.

To further verify that the observed broadening is not caused by photometric errors, we conducted Monte Carlo simulations of a synthetic cluster with a single stellar population. Our goal is to simulate the most populated red MS and verify that a single population cannot reproduce also the blue sequence observed in the CMD. To do so,
we randomly drew star masses using a Salpeter mass function, normalizing to the observed (completeness-corrected) number of stars. To each mass we associated the corresponding magnitude and color from the best-fit isochrone (derived as described in Sect.~\ref{s:iso}) and we shift the intrinsic color of the model to match the fiducial line of the red MS (derived as described in Sect.~\ref{s:fiducial}). A component of unresolved binary stars, derived from the same mass function, is added to a fraction of the sampled stars. We use a flat distribution of primary-to-secondary mass ratios and we adopt a binary fraction of $\simeq$ 15\% \citep{davi+08}. Finally, we add the photometric errors with a distribution derived from the artificial star tests. 
The derived simulated CMD (blue dots) and the color distribution below the MSK (blue lines) are superimposed in Fig.~\ref{f:cmdbin1} to the observed CMD and observed color distribution.   

Our analysis indicates that NGC\,6397 hosts a double MS below the MSK, as observed in NGC\,2808 and M\,4. In particular, the ``primordial' population (i.e., the blue MS) is the less populated one, similarly to what is observed in M\,4 and in agreement with the results obtained by \citet{milo+12a}.

\section{Deriving the Fiducial Line}
\label{s:fiducial}

To obtain the NGC\,6397 fiducial line, we follow the same approach used in C16 for the cluster M\,4, where the primordial (blue) population is not the predominant one. 

Briefly, we used a kernel-density-estimation (KDE) to estimate the 2D probability density function (PDF) underlying the data in the CMD. We found the PDF global maximum and by moving along the direction of minimum gradient, we traced the ridge line. Then, to ensure the smoothness of the ridge line, we computed the corresponding bezier curve and we used this smooth approximation as the final fiducial line. As demonstrated in C16, this method is not prone to biases due to the presence of binaries and it accurately traces the fiducial line of the most prominent population when multiple MS are present. 

\begin{figure}[!thp]
\includegraphics[width=1\columnwidth]{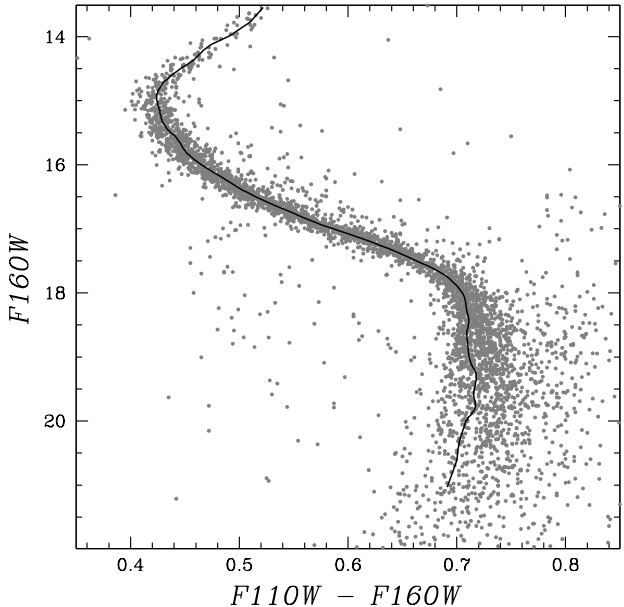}
\caption{$F160W$ vs $F160W - F110W$ NGC\,6397 CMD with the superimposed fiducial line (black line) derived as described in Sect.~\ref{s:fiducial}.}
\label{f:cmd_rl}
\end{figure}

Next, we adopted the following empirical approach, specifically customized for the case in which the MS that we want to trace is the less populated. We applied a magnitude-color cut in the CMD and selected stars fainter than $F160W \sim 18.5$ mag and redder than $(F110W - F160W)_{fiducial}$, where $(F110W - F160W)_{fiducial}$ is the color of the fiducial line that traces the most prominent population. After removing these objects from the catalog, we re-derived the fiducial line and its error on the new dataset.  

Finally, the actual value for the fiducial line and its uncertainty is derived adopting a bootstrap approach. We drew one thousand samples from the cluster catalog (after the magnitude-color cut), using the same total number of stars, and allowing for repetitions. The fiducial line for each bootstrap catalog is calculated using the procedure outlined above. We then divided the ensemble of one thousand fiducial realization into 0.05 mag bins in $F160W$. We estimated the fiducial value and its error, assuming the bin center as the $F160W$ fiducial value and half the bin size as its error. For the $F110W - F160W$ color, we adopted the mean and standard deviation over the ensemble of fiducials, using the points that fell within each magnitude bin. 

\begin{figure}[!thp]
\includegraphics[width=1\columnwidth]{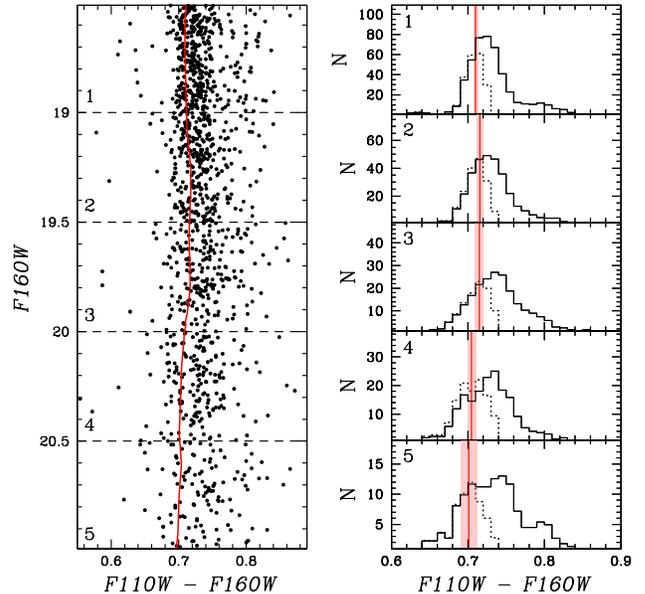}
\caption{Similar to Figure~\ref{f:cmdbin1} but with the superimposed derived fiducial line. For each magnitude interval we report the fiducial line (red line) and its mean error (light red region). The dotted histograms represent the color distribution of stars in each specific bin after we applied the magnitude-color cut described in Sect.~\ref{s:fiducial}.}
\label{f:cmdbin2}
\end{figure}

Figure~\ref{f:cmd_rl} shows the cluster CMD with the superimposed fiducial line obtained as described above, whereas in Figure~\ref{f:cmdbin2} we show the same zoom-in of the region below the MSK (as in Figure~\ref{f:cmdbin1}) and the corresponding color distribution in the five magnitude bins, to better illustrate the shape of the new fiducial line in this portion of the CMD. The mean fiducial color (red line) and its mean error (light red region) are reported in each bin. This comparison demonstrates that the newly derived fiducial line reproduces quite well the peak of the blue population.

To test the robustness of the obtained measure we performed the following tests: we iterated the procedure by applying different magnitude cuts through a $\pm 0.25$ mag step with respect to the default value quoted above (i.e, $F160W = 18.50$). Then, we verified that the derived fiducial lines are consistent with each other within the errors and that the shape is not altered by the different selection choices. Furthermore, for each bootstrap catalog we derived the mean color in the five magnitude bins of Fig.~\ref{f:cmdbin1}, applying a smoothed naive estimator \citep{silv86} to be insensitive to a particular binning starting point. To estimate the final fiducial magnitude and color values, we adopted the same procedure described above (i.e., the fiducial magnitude $F160W$ is the bin center and the fiducial color is the mean over the different realization). As pointed out in C16 for the cluster M\,4, due to the empirical procedure applied to derive the fiducial line, we acknowledge that the uncertainties derived in Sect.~\ref{s:prob} could be slightly underestimated.

\section{Constraints on the GC parameters} 

\subsection{Isochrones fitting: deriving the GC age}
\label{s:iso}
To derive the cluster best-fit parameters in terms of age, metallicity, distance and reddening, we followed the same approach as in C16.  We compared the fiducial line with a set of stellar models from the Victoria-Regina evolutionary code \citep{vand+14}. We transformed the models to the WFC3/IR filter system by means of synthetic photometry using the MARCS library of stellar spectra \citep{gust+08} and the most updated WFC3/IR throughputs and zero points. To take into account the extinction, we applied the \cite{fitz99} extinction law to the spectra, before integrating them under the throughput curves. 

\begin{table*}[!thb]
\begin{center}
\caption{GC prior parameters and ranges}
\begin{tabular}{ccccccccc}
\hline
\hline
Age$_{range}$ & [Fe/H] & $\Delta$ [Fe/H] & ref. & $(m-M)_0$ & $\Delta$ $(m-M)_0$ & E\,(B-V) & E\,(B-V)$_{range}$ &ref.\\
 (1) & (2) & (3) & (4) & (5) & (6) & (7) & (8) & (9)\\
\hline
 & & & & & & & &\\ 
11.0\,--\,15.0  & -2.02  & $\pm$ 0.20 & 1 & 12.12 &  $\pm$ 0.25 & 0.18 & $\pm$ 0.10 & 2\\
 & & & & & & & & \\ 
\hline
\end{tabular}
\tablecomments{Columns (1): Age interval (Gyr). (2): Metallicity reference value (dex). (3): Metallicity interval (dex). (5): Reference for the metallicity value ((4) \citet{kraiva03}. (5): Distance modulus reference value (mag). (6): Distance modulus interval (mag). (7): Reddening reference value (mag). (8): Reddening interval (mag). (9) Reference for distance modulus and reddening values. ( (2) \citet{reigiz98}).}
\label{t:GC_prior}
\end{center}
\end{table*} 

We then built a grid of isochrones over a large range of the parameter space, allowing age, metallicity, distance and reddening to vary freely over reasonable uniform priors that are consistent with the literature. The reference value and the adopted intervals are reported in Table~\ref{t:GC_prior}. In detail, we used a fixed age range (i.e, from 11 to 15 Gyr), an interval of $\pm$ 0.20 dex for the metallicity, and an interval of $\pm$ 0.20 and 0.10 mag for distance and reddening, respectively. During the analysis, we iteratively readjusted the parameter range in order to center the derived best-fit results in the new interval and to ensure that the marginal probability in each variable falls to zero well within the ranges. The grid is obtained by adopting steps of 0.1 Gyr for the age, 0.01 dex for the metallicity, and 0.01 mag for distance and reddening. 

As in C16, the only a priori fixed parameters are the the [$\alpha$/Fe] ratio, the helium abundance, and the extinction coefficient $R_V$. For the [$\alpha$/Fe] ratio we adopted the value [$\alpha$/Fe] = +0.4 dex (Spectroscopic measures indicate [$\alpha$/Fe] = +0.36 dex, Carretta et al. 2010). The original helium abundance in the stellar models was obtained using Eq.~1 and Eq.~2 from \citet{genn+10}, after the ratio between [Fe/H] and [M/H] was taken into account using Eq.~3 from \citet{sala+93}. We assumed the solar mixture from \citet{aspl+09}, a primordial helium abundance of $Y_P = 0.2485$ from \citet{izot+07} and \citet{peim+07}, and a ratio of $\Delta Y / \Delta Z$ = 1.5. Finally, for the extinction coefficient, $R_V$, we adopted the value $R_V = 3.1$. 

\begin{figure}[thp]
\includegraphics[width=1\columnwidth]{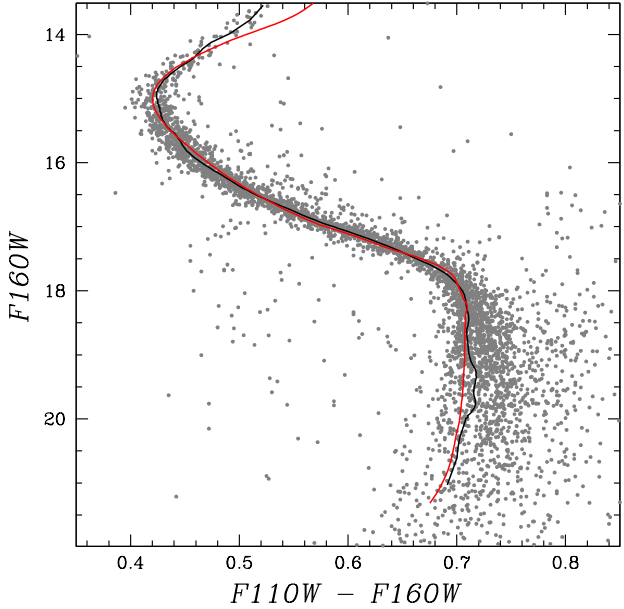}
\caption{$F160W$ vs $F160W - F110W$ NGC\,6397 CMD with superimposed the fiducial line (black line) and the best-fit isochrone (red line) derived as described in Sect.~\ref{s:iso}. Best-fit parameters are reported in Table~\ref{t:GC_param}.}
\label{f:cmd_iso}
\end{figure}

The best-fit isochrone has been obtained comparing the fiducial line with the isochrone grid and deriving for each isochrone the posterior PDF. Due to choice of uniform priors in our parameters, the posterior PDF is proportional to the likelihood $\mathcal{L}$, which is calculated using the following equation:
\begin{equation}
\mathcal{L} \simeq  \exp\,(-\frac{1}{2} \chi^{2})
\label{e:like}
\end{equation}
where the term $\chi^2$ in Eq.~\ref{e:like} is defined as:
\begin{equation}
\chi^2 = \sum_{i=1}^{N}\frac{(\Delta col_{i})^{2}}{\sigma^{2}_{i}}
\label{e:chi}
\end{equation}
where $\Delta col$ is the difference in color between the isochrones and the fiducial (i.e. $(F110W - F160W)_{iso} - (F110W - F160W)_{fiducial}$), calculated at each point in the fiducial line. $\sigma$ is the error associated with the fiducial color points, derived as described in Sect.~\ref{s:cmd}. The best-fit isochrone is the one that maximizes the joint PDF for the four parameters.

\begin{figure}[!thp]
\includegraphics[width=1\columnwidth]{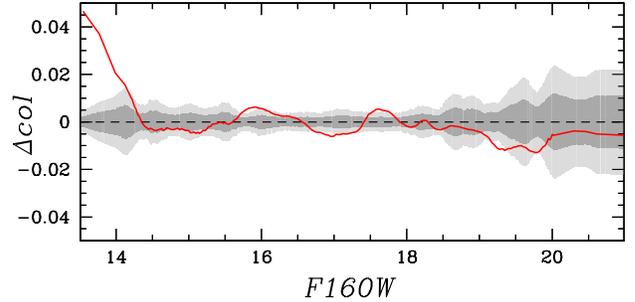}
\vspace{-4truecm}
\caption{Residual of the fit between the best-fit isochrone and the fifducial line. Dark and Light gray region represent the 1$\sigma$ and 2$\sigma$ fiducial color errors.}
\label{f:residual}
\end{figure}

Fig.~\ref{f:cmd_iso} shows the GC CMD with superimposed the fiducial line (black line) and the best-fit isochrone (red line). The derived parameters and uncertainties from the best-fit isochrone (obtained as described in Sect.~\ref{s:prob}) are reported in Table~\ref{t:GC_param}. The fit residuals (i.e., the term $\Delta col$ from Eq.~\ref{e:chi}) as a function of the magnitude $F160W$ are shown in Fig.~\ref{f:residual}. The best-fit isochrone reproduces the fiducial line quite well, with some discrepancies in the sub-giant branch (SGB) region and below the MSK. For the bright portion of the CMD, a similar trend was observed also in C16, where all of the isochrones were slightly redder ($\Delta col \sim$ 0.02\,--\,0.04 mag) than the fiducial line in the SGB and red giant branch region. We note that the same behavior was pointed out also in \citet{vand+13} in their analysis of visible-light photometry. This difference can be caused by a series of factors: small zero point or systematic errors in the color-$T_{eff}$ relations, problems in the stellar models concerning diffusion, convection, and/or the atmospheric boundary conditions \citep[see][for a detailed discusion]{vand+13}. Similar discrepancies between models and observations in the red giant branch, over a broad metallicity baseline, were found also by \citep{cohen+15} from the analysis of wide field ground-based IR ($J - K_{s}$) photometry of 12 Galactic GCs.

\begin{table*}[!thb]
\begin{center}
\caption{GC best-fit parameters}
\begin{tabular}{cc|cc|cc|cc}
\hline
\hline
Age & $\sigma$ & [Fe/H] & $\sigma$ & $(m-M)_0$ & $\sigma$ & E\,(B-V) & $\sigma$ \\
 (1) & (2) & (3) & (4) & (5) & (6) & (7) & (8) \\
\hline
 & & & & & & & \\
 12.6  & $^{+0.6}_{-0.7}$ & -1.88 & $^{+0.04}_{-0.04}$ & 12.02 & $^{+0.03}_{-0.03}$ & 0.22 &  $^{+0.015}_{-0.015}$\\
 & & & & & & & \\
\hline
\end{tabular}
\tablecomments{Columns (1): Age (Gyr). (2):  Age uncertainty (Gyr, 68\% confidence interval). (3): Metallicity (dex). (4): Metallicity uncertainty (dex, 68\% confidence interval). (5) Distance modulus (mag). (6) Distance modulus uncertainty (mag, 68\% confidence interval). (7): Reddening (mag). (8) Reddening uncertainty (mag, 68\% confidence interval).}
\label{t:GC_param}
\end{center}
\end{table*} 

\begin{figure*}[!thp]
\begin{center}
\includegraphics[scale = 0.6]{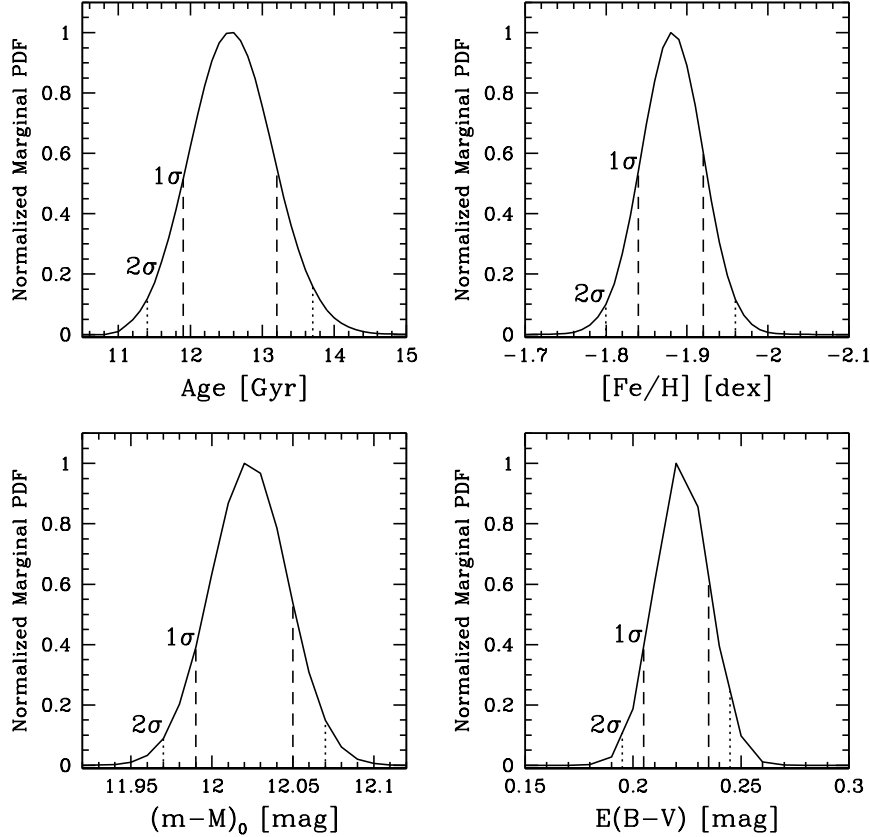}
\caption{1D posterior PDF for the 4 parameters (top left: age, top right: metallicity, bottom left: distance modulus, and bottom right: reddening). Each 1D PDF is obtained by the marginalization of the 4D PDF over the other three parameters. 1$\sigma$ (dashed lines) and 2$\sigma$ (dotted lines) confidence intervals, defined as described in Sect.~\ref{s:prob}, are also shown. 1D posterior PDF have been normalized by setting each PDF peak = 1.}
\label{f:prob1D}
\end{center}
\end{figure*}

\begin{figure*}[thp]
\hspace{-0.1cm}
\includegraphics[scale=0.3]{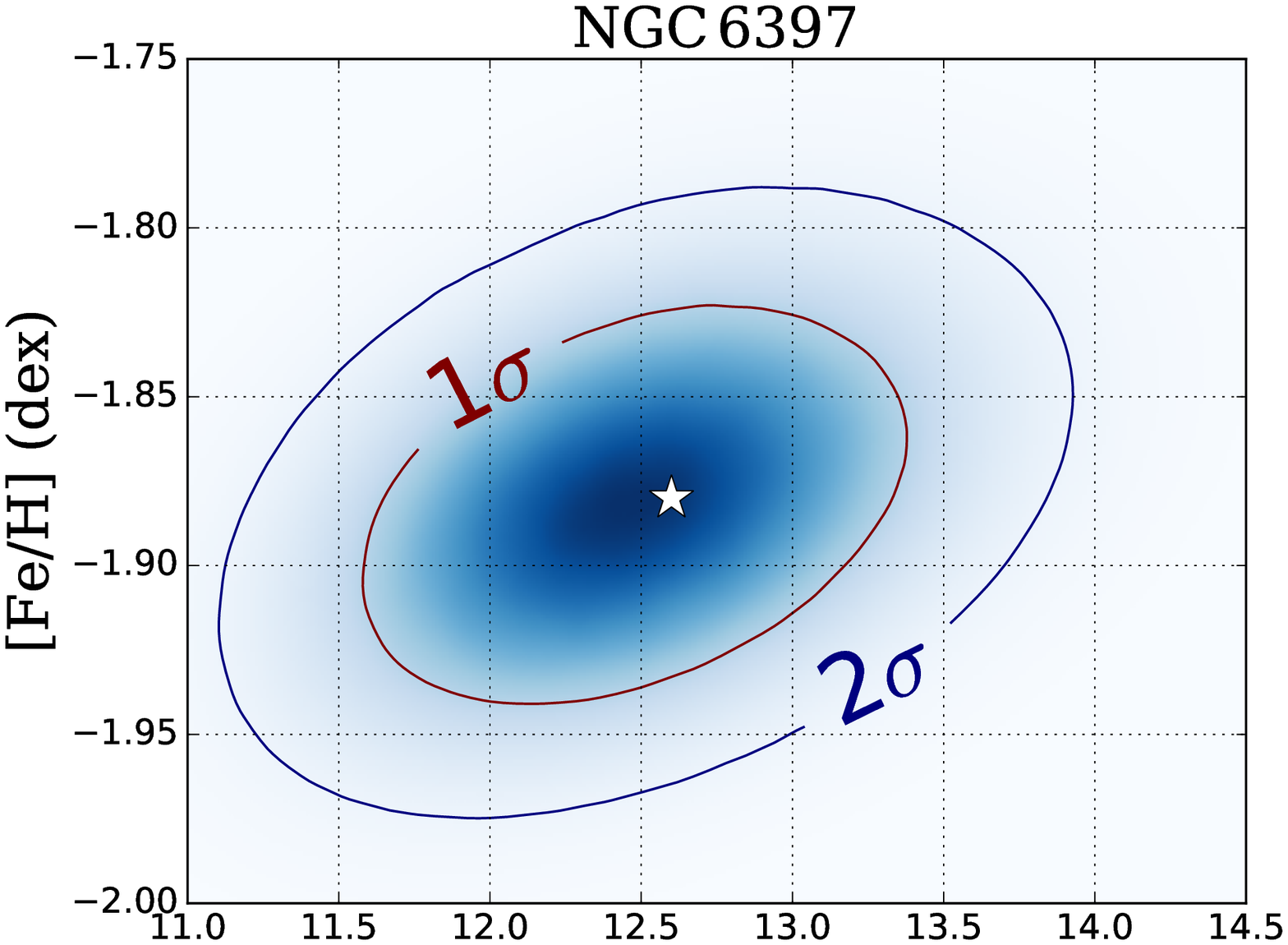}
\hspace{-0.25cm}
\includegraphics[scale=0.3]{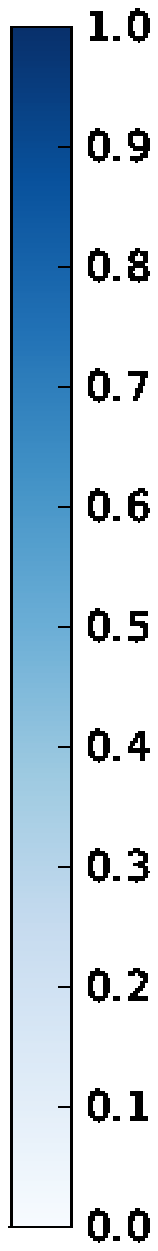}\\
\hspace{-0.1cm}
\includegraphics[scale=0.3]{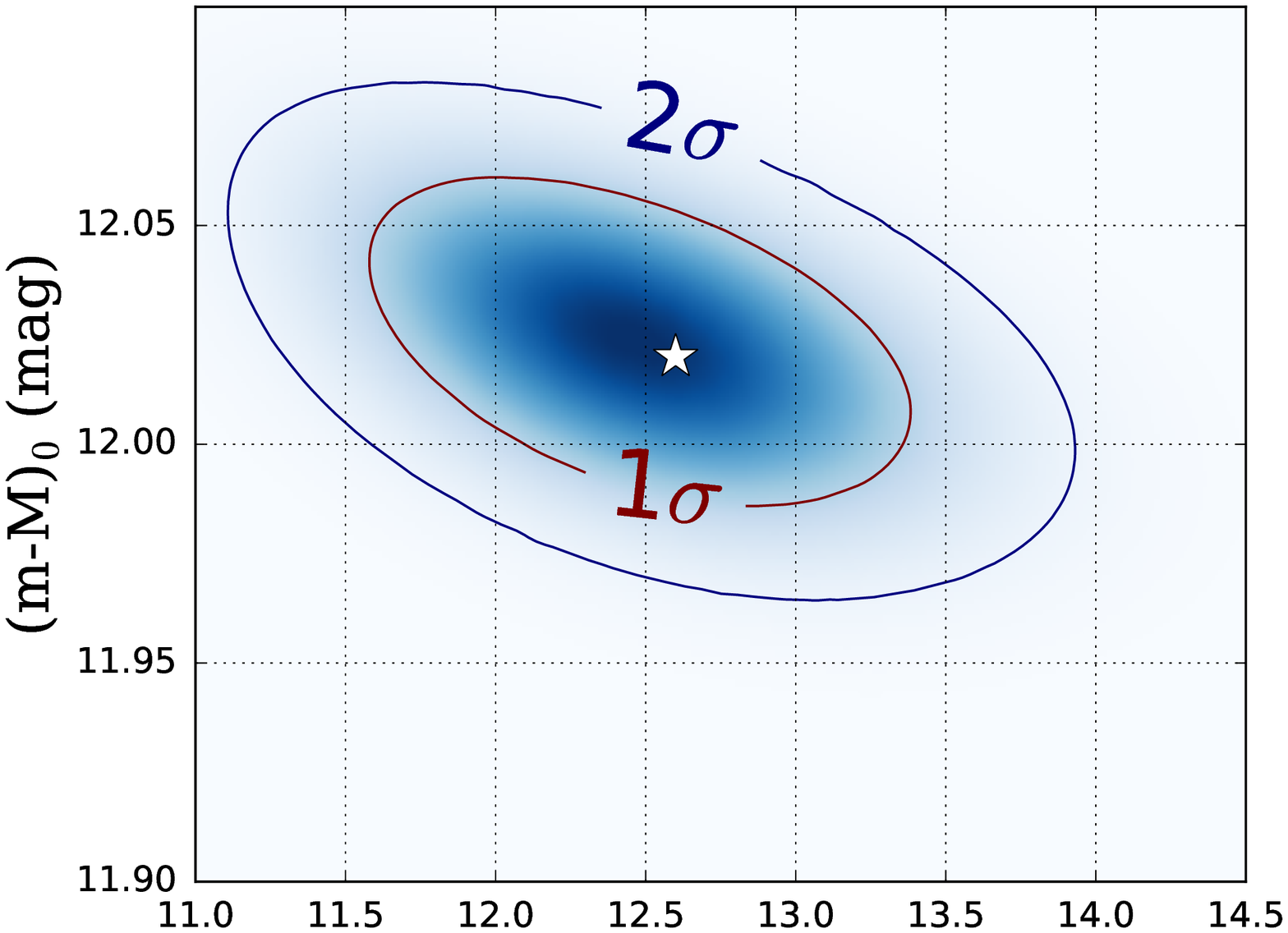}
\hspace{-0.15cm}
\includegraphics[scale=0.3]{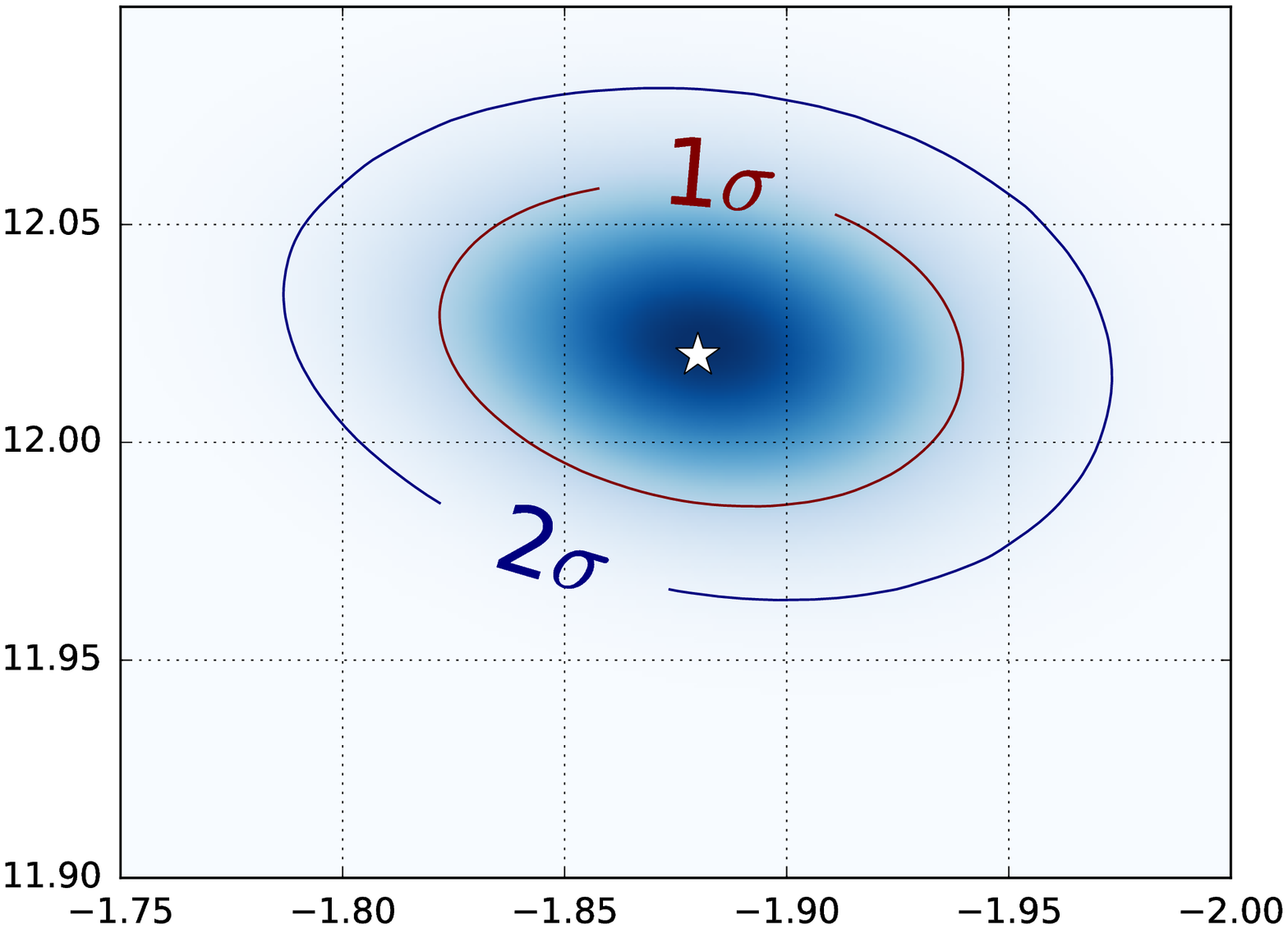}\\
\hspace{-0.25cm}
\includegraphics[scale=0.3]{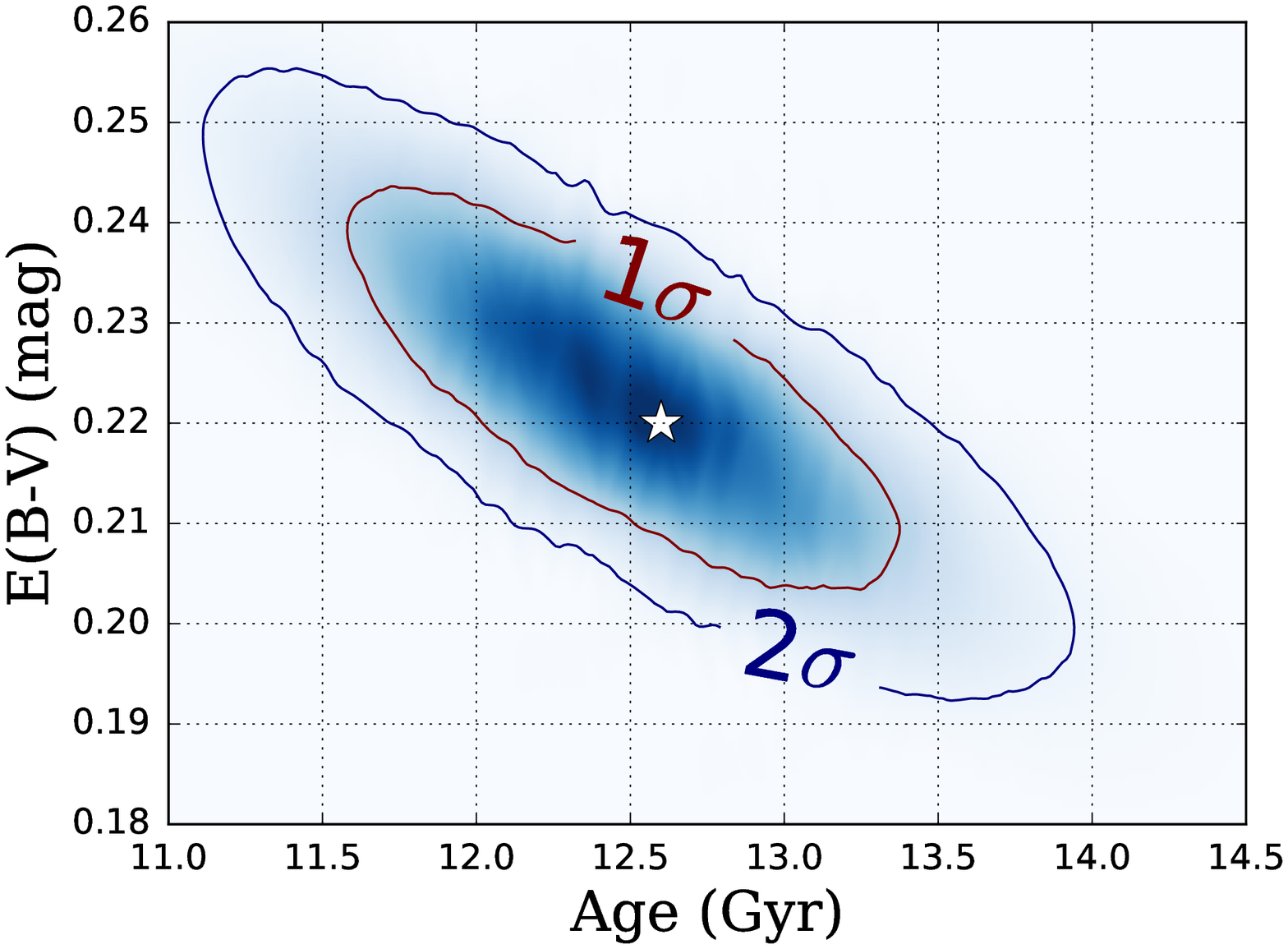}
\hspace{-0.15cm}
\includegraphics[scale=0.3]{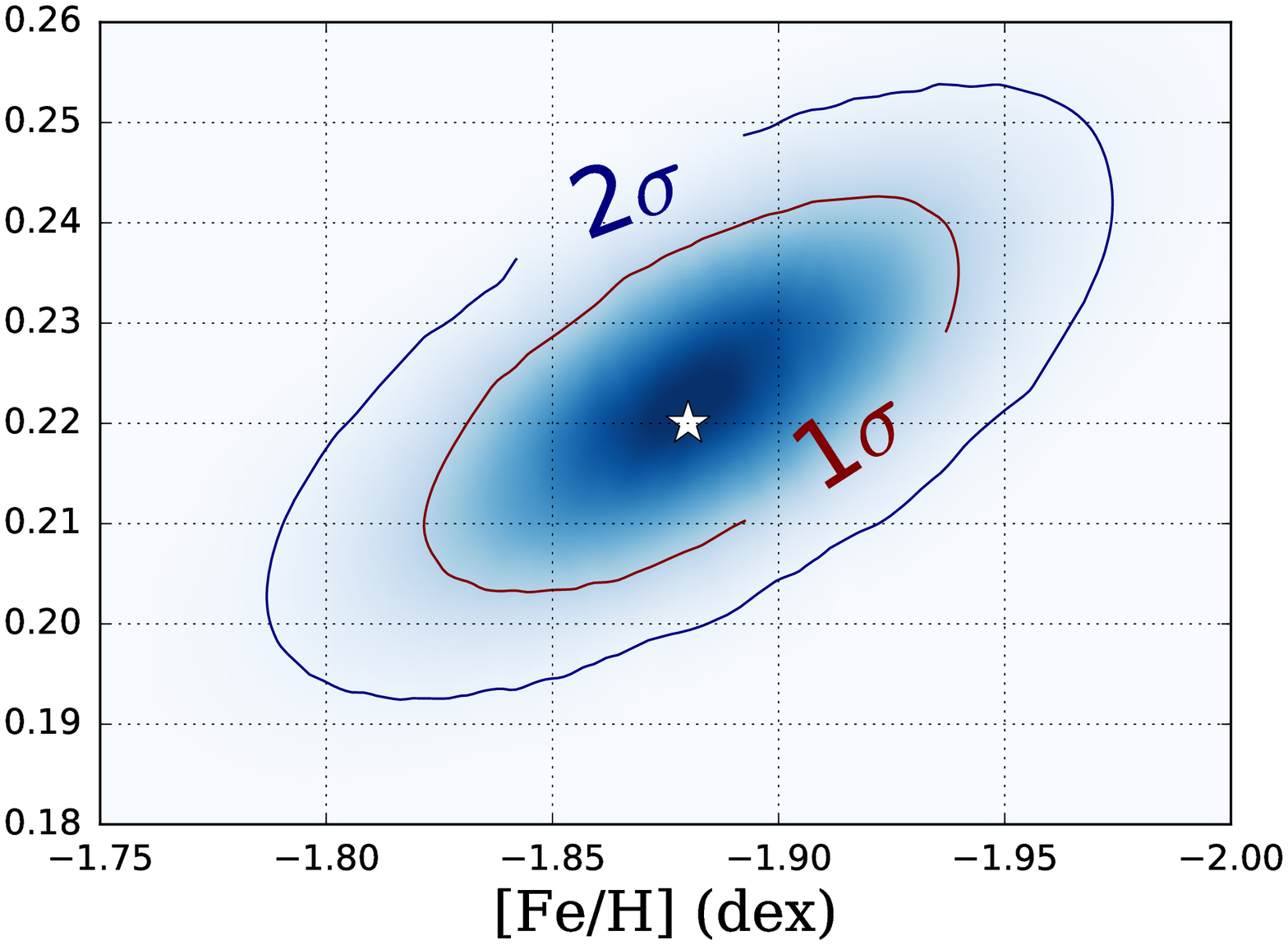}
\hspace{-0.15cm}
\includegraphics[scale=0.3]{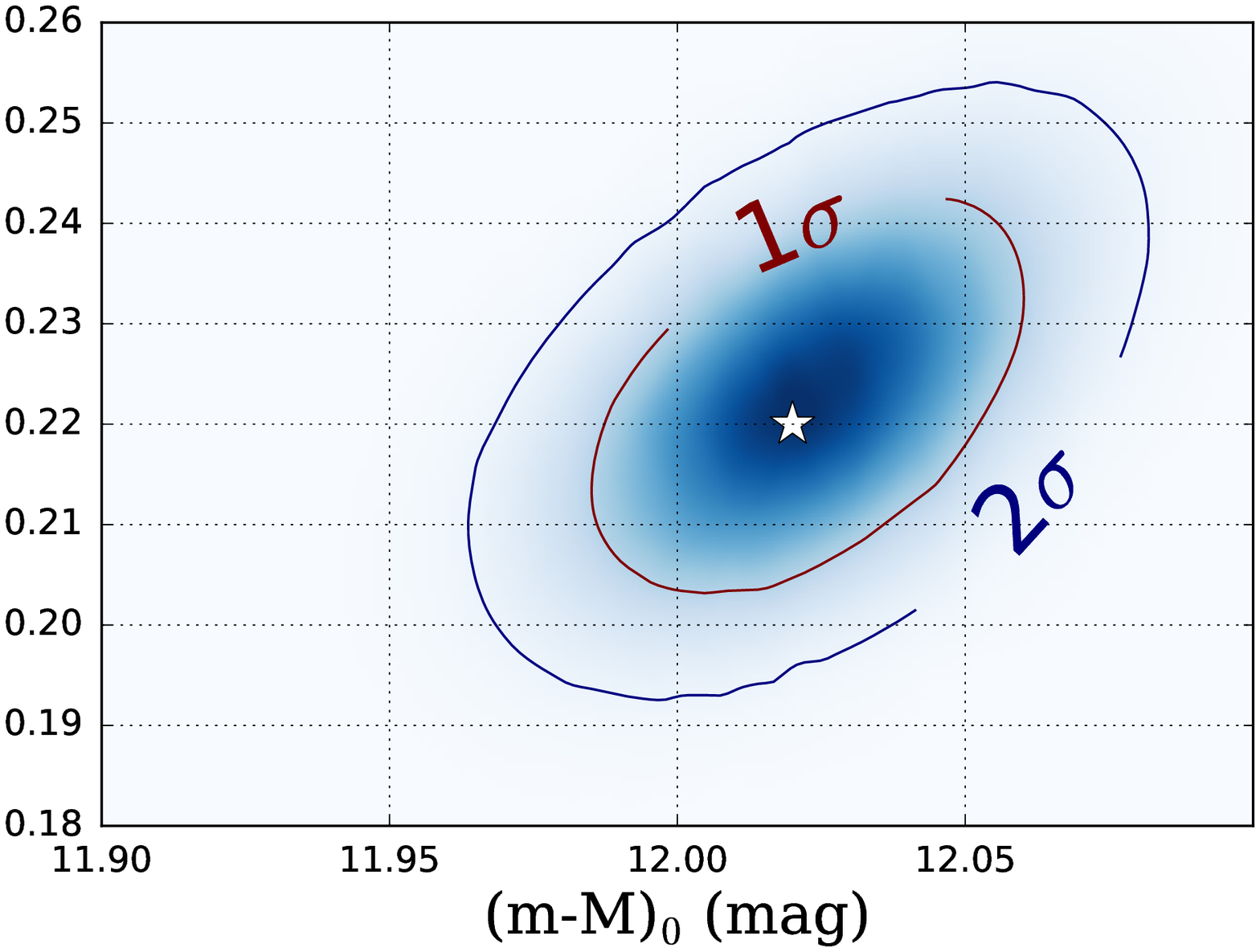}
\caption{2D posterior PDFs of all of the parameter combinations for NGC\,6397 (upper panel: age vs metallicity 2D PDF; middle panels: age vs distance modulus, metallicity vs distance modulus 2D PDFs; lower panels: age vs reddening, metallicity vs reddening, and distance modulus vs reddening 2D PDFs). The 1$\sigma$ (red lines) and 2$\sigma$ (blue lines) regions are defined as the smallest regions such that the integral of the 2D PDFs within the regions are equal to 0.68 and 0.95. Color codes for the 2D PDFs are shown in the upper right sub-panel. The white stars indicate the values for which the 4D PDF has a maximum.}
\label{f:prob2D}
\end{figure*}

Below the MSK the fiducial line exhibits a change with respect to the ``expected'' color trend, bending slighlty towards the red at $F160W \sim$ 19 mag, before turning back towards bluer colors at $F160W \sim$ 20 mag, producing a step in the fiducial shape which is not present in the isochrone. The low density of stars in this region of the CMD, coupled with the intrinsic characteristic of the method adopted to derive the fiducal line, can produce this deviation from a smooth shape for the fiducial. To verify if the results obtained are influenced by this particular shape of the fiducial line, we derived the best-fit parameters using the fiducial line obtained from the mean color in the five magnitude bins below the MSK (i.e, using only five data point below $F110W$ = 18.50 mag). This method is less prone to small scale variations and attribute less weight to this portion of the CMD when the fiducial line is compared with the isochrones grid. Adopting this fiducial, we obtained the following values for the best-fit parameters: age = 12.80 Gyr, metallicity [Fe/H] = -1.87 dex, distance modulus $(m-M)_0$ = 12.01 mag, and reddening E(B-V) = 0.21 mag.

Finally, we also derived the best-fit parameters using the fiducial lines obtained with the different magnitude cuts (i.e., $\pm 0.25$ mag step with respect to the default value $F160W =$ 18.50 mag). In this case, we obtained variations of $\pm$ 0.2 Gyr for the age, $\pm$ 0.02 dex for the metallicity, $\pm$ 0.01 mag for the distance modulus, and $\pm$ 0.01 mag for the reddening. 

\subsection{Comparison with literature estimates}
\label{s:comp_lit}

To test that the isochrone fitting procedure provides reasonable values for the best-fit parameters, we compared our results with several literature estimates. Due to its proximity, NGC\,6397 has been thoroughly observed and analyzed in many studies which derived its age using different methods and tracers. In this context, large discrepancies exist in the age determination, with values ranging from 11.47 Gyr \citep{hans+07} to 13.9 Gyr \citep{gratt+03}. A nice summary of the main literature age estimates is provided in Table~2 in \citet{torr+15}. For the metallicity, spectroscopic measures in the literature yield [Fe/H] $\simeq$ -2.0 ([Fe/H] = -2.03 $\pm 0.05$ dex, Gratton et al. 2003, [Fe/H] = -2.02 $\pm 0.07$ dex, Kraft \& Ivans 2003, [Fe/H] = -1.99 dex, Carretta et al. 2009, and [Fe/H] = -2.10 dex, Husser et al. 2016, and references therein). From the analysis of the GC white dwarf cooling sequence, \citet{hans+13} derived [Fe/H] = -1.8 dex, while \citet{anto00}, from the analysis of the GC CMD using Stromgren photometry, obtained [Fe/H] = -1.84 dex. As in C16, the most direct comparison is with the results obtained in \citet{vand+13}, which using the same stellar evolutionary code, although with slightly different prescriptions for the isochrone computation, derived the cluster best-fit parameter from the analysis of HST ACS visible-light photometry. \citet{vand+13} derived an age of 13.0 Gyr, assuming a metallicity [Fe/H] = -1.99 dex. The age estimate is in good agreement with the value obtained in this work, while our metallicity is slightly higher, although consistent, within 2$\sigma$, with the literature values. For the true distance modulus, literature values span the interval 12.03\,--\,12.13 mag \citep{gratt+03, reigiz98,hans+07}, whereas the reddening is E(B-V) = 0.18 $\pm$ 0.01 mag \citep[][and references therein]{reigiz98, rich+08}. While the derived value for the distance modulus is in good agreement with previous works, our estimate of the reddening E(B-V) is slightly higher (of the order of 0.02\,--\,0.04 mag). However, literature values translate to an apparent distance modulus $(m-M)_V$ = 12.59\,--\,12.68 mag, in agreement within the errors with our estimate ($(m-M)_V$ = 12.70 mag).    

Finally, during the review process, \citet{brown+18} published a paper in which they obtained a direct trigonometric parallax of NGC\,6397, exploiting the HST/WFC3 spatial scanning mode. The authors obtained a true distance modulus $(m-M)_0$ = 11.89 $\pm$ 0.07 $\pm$ 0.09 mag and an absolute cluster age of 13.4 $\pm$ 0.7 $\pm$ 1.2 Gyr. Their distance modulus is shorter than our estimate (and literature estimates) at a level of 1$\sigma$\,--\,2$\sigma$, altough consistent with recent dynamical analysis \citep{watk+15}. Despite using the same stellar evolutionary code, the age estimate is slightly higher than the one derived by \citet{vand+13} and by this work (+0.4 and +0.8 Gyr, respectively), although consistent within 1$\sigma$ with both works.

\subsection{Posterior probability density function:\\ deriving age uncertainties}
\label{s:prob}

To measure the uncertainties for each best-fit parameter we derived, for each isochrone of the grid, the joint posterior PDF. The latter, due to choice of uniform priors for our parameters, is proportional, within the specified prior ranges, to the likelihood  $\mathcal{L}$, obtained from Eq.~\ref{e:like} and Eq.~\ref{e:chi}. Then, to derive the 1D posterior PDF for each parameter (1D P(X)), we marginalized the 4D PDF ($P(X,Y,W,Z)$ where X, Y, W, Z are age, metallicity, reddening, and distance) over the remaining three parameters (i.e., Y, W, Z). Each individual uncertainty is obtained from the cumulative distribution of the marginalized 1D distribution. 1$\sigma$ (2$\sigma$) confidence intervals are defined as the area enclosed within 16\% (2.5\%) and 84\% (97.5\%) of each cumulative distribution. This would correspond to the {\it true} 1$\sigma$ and 2$\sigma$ uncertainties if the 1D PDFs were Gaussian. Fig.~\ref{f:prob1D} shows the derived random uncertainties for the 4 parameters. In particular, we obtain that random age uncertainties are of the order of $\sigma \sim$ 0.7 Gyr.

Considering the uncertainties in the best-fit parameters when a different fiducial line is adopted (i.e., when a different magnitude cut is applied or a different method for tracing the faint part below the MSK), we acknowledge that the derived age uncertainties could be underestimated. Taking into account the age range derived in Sect~\ref{s:iso} and considering the uncertainties in the method, we estimate that the value $\sigma \sim$ 1.0 is more representative of the GC age uncertainty.   

To illustrate the correlations and interdependencies between the various parameters, we show in Figure~\ref{f:prob2D} the 2D posterior PDFs of all the parameters. These 2D PDFs are obtained by the marginalization of the 4D PDF over two parameters instead of three. 

Finally, we note that the quoted uncertainties only take into account random noise (number statistics and measurement errors). We acknowledge that the total uncertainty on GC parameters is further increased by the presence of a systematic component due to e.g. the choice of the stellar evolution library, stellar atmospheres and possible zero-point offsets (as well as other, more subtle, possibly unknown sources of uncertainty). Moreover, it is important to keep in mind that the results obtained in this study, as well as in C16, are strictly model-dependent. Adopting a different set of stellar models can produce different results in terms of best-fit parameters \citep[see for a detailed discussion][]{sarac+18}. However, such issues are common for most age derivations of GCs. Our results demonstrate that using the IR photometry and the MSK as an age diagnostic represent a new venue to determine GCs age. 

\section{Age-metallicity Relation}
\label{s:age_met}

Adding NGC\,6397 to the C16 sample allow us to provide a first estimate of the age-metallicity relation obtained from IR photometry. Although the sample is limited to five GCs, it spans a significant range of metallicity (from [Fe/H] $\sim$ -1.9 dex to [Fe/H] $\sim$ -0.7 dex). The slope of this relation is of fundamental importance to test high-resolution N-body simulation of galaxy formation, and inform the mass-merger history that leads to the build up of Milky Way-type halos \citep{macgil04,font+11,beer+12}.  

\begin{figure}[!thp]
\includegraphics[width=1\columnwidth]{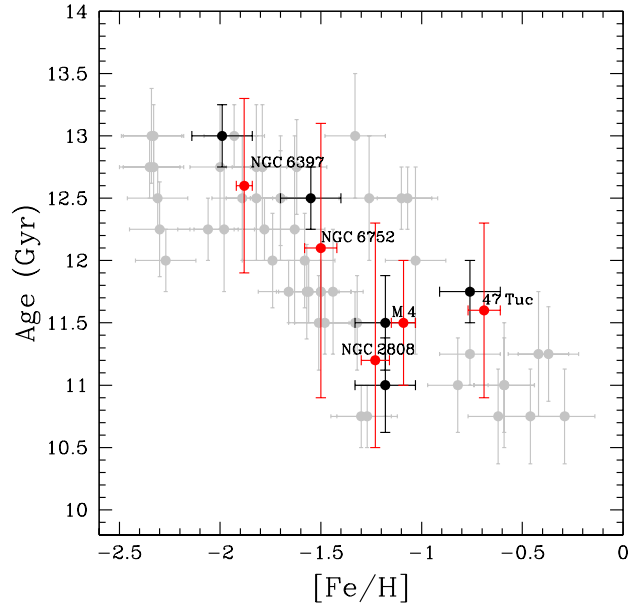}
\caption{Age-[Fe/H] relation derived in the present work (red dots) compared with \citet{vand+13} relation for the clusters in common (black dots). As a reference, we reported also the age-[Fe/H] values for the whole \citet{vand+13} sample (grey dots).}
\label{f:agemet}
\end{figure}

Fig.~\ref{f:agemet} shows the age-metallicity relation derived using the five GCs in our sample (red dots), superimposed on the age-metallicity relation derived by \citet{vand+13} from their analysis (grey dots). \citet{vand+13} measures for the clusters in common are reported as black dots. The two age-metallicity relations are in good agreement, with a similar slope and normalization. We note that age uncertainties in \citet{vand+13} have been derived using a different approach with respect to our method. They are defined as the difference between the best-fit ages obtained by the different authors that independently determined the GCs distance moduli and ages. Hence, the error arising from the fitting of the horizontal branches and isochrones is $\pm 0.25$ Gyr. When uncertainties associated to the GC distances and chemical compositions are taken into account, the error is of the order of $\sim \pm$ 1.5\,--\,2.0 Gyr.

The results from this project can be greatly enhanced with next generation space IR telescopes, such as the {\it James Webb Space Telescope} ({\it JWST}).  {\it JWST} offers increased photometric sensitivity, redder wavelength coverage, higher resolution, and larger fields of view compared to {\it HST}/WFC3-IR.  This will enable high-precision detection and characterization of the MSK in stellar populations out to larger distances (e.g., clusters over a range of ages and metallicities) and towards extincted sightlines such as the Galactic bulge. Such data will yield new constraints on the age-metallicity relation of stars.  Additionally, {\it JWST} can easily measure this feature in all Milky Way dwarf satellites and greatly enhance studies of their ancient star formation histories.

\section{Summary and Conclusion}
\label{s:summary}

In this study, we analyzed new, deep, {\it HST} WFC3/IR observations of the old metal-poor GC NGC\,6397. Following the work presented in C16, we exploited the constraints imposed on GC properties by leveraging the shape of the MS color-magnitude relation in pure IR CMDs. 

We discovered the presence of at least two MS, being the bluer, less-populated MS, the one corresponding to the first stellar generation. We used an ad-hoc fitting method to derive the fiducial line of the cluster and we compared it with a grid of isochrones over a large range of parameter space. Age, metallicity, distance, and reddening can vary freely within reasonably selected ranges. From the comparison between the fiducial line and the isochrones, we calculated the joint posterior 4D PDF and we derived the best-fit isochrone by maximizing the 4D PDF. We measure a cluster age of 12.6 Gyr, in reasonable agreement with the results obtained by \citet{vand+13}, who used the same stellar evolution code to analyze visible-light ACS photometry of NGC\,6397. 

Age uncertainty is derived by calculating the 1D posterior PDF. This involves marginalizing the 4D PDF over metallicity, distance, and reddening. The random age uncertainty is of the order of $\sim \sigma$ 0.7 Gyr. Taking into account the different values for the GC parameters obtained with the different fiducials and the uncertainties in the method, we estimate that an uncertainty of $ \sigma \sim$ 1.0 Gyr is more representative of the true random uncertainty. 


Our results confirm that the IR color-magnitude relation and the MSK in the lower MS represents a promising tool to obtain absolute ages of GCs with sub-Gyr accuracy.


\end{document}